\documentclass[12pt]{article}

\begin{document}
\begin{titlepage}
\title{\vskip -70pt
\begin{flushright}
{\normalsize \ DAMTP-1998-80}\\
\end{flushright}
\vskip 20pt
{\bf Conservation Laws in a First Order Dynamical System of Vortices }
}
\vspace{1cm}
\author{{N. S. Manton}\thanks{e-mail: N.S.Manton@damtp.cam.ac.uk}
\hspace{.2cm}  and
\hspace{.2cm}
 {S. M. Nasir}\thanks{e-mail: S.M.Nasir@damtp.cam.ac.uk}\\
{\sl Department of Applied Mathematics and Theoretical Physics}\\
{\sl University of Cambridge} \\
{\sl Silver Street, Cambridge CB3 9EW, England}}
\date{September, 1998}
\maketitle
\thispagestyle{empty}
\vspace{1cm}
\begin{abstract}
\noindent
Gauge invariant conservation laws for the linear and 
angular momenta are studied
in a certain 2+1 dimensional first order  dynamical
model of vortices in superconductivity. In analogy with fluid vortices
it is possible to express the linear and angular momenta as low moments
of vorticity. The conservation laws are compared with those obtained
in the moduli space approximation for vortex dynamics.

\vspace{.5cm}

\end{abstract}
\end{titlepage}

\section{Introduction}

Recently, a 2+1 dimensional field theory was proposed to describe
the non-dissipative dynamics of
vortices in thin-film superconductors \cite{Man1}.  
The model, which has $U(1)$ gauge invariance, has a
Lagrangian of Schr\"{o}dinger--Chern--Simons type
containing terms linear in the first time
derivatives of the fields, and for static fields, the Lagrangian reduces
to the standard Ginzburg--Landau model. Interestingly, in this model 
two vortices can be shown to orbit around each
other. It has been argued that such motion occurs in superconductors
at very low temperature \cite{Ait, Sto}.

For certain values of the coupling
constants, the model has a large space of static solutions consisting of
multi-vortices obeying Bogomol'nyi equations \cite{Bog}. The space of such
$N$-vortex solutions, whose parameters are the vortex positions, is
known as the $N$-vortex moduli space. When the coupling constants have
slightly different values, one can study vortex dynamics in the model 
by obtaining a reduced theory, where one projects the
motion onto the moduli space of Bogomol'nyi vortices. This is the
essence of the moduli space approximation to describe soliton dynamics
\cite{Man2}. The dynamical variables of the reduced theory are just
the time-varying vortex positions. The moduli space approach cannot
deal with the dynamics of interacting vortices and anti-vortices.

The moduli space approximation has been established rigorously by Stuart
\cite{Stu} in the context of slowly moving Bogomol'nyi vortices in the
relativistic Abelian Higgs model. However, in the
proposed Schr\"{o}dinger--Chern--Simons type of model of vortices it is
not yet certain that one can use the moduli space approximation 
faithfully in order to extract dynamics. Obtaining conservation laws
provides an important consistency check.

The main conclusion in \cite{Man1} was the Lagrangian of the reduced
theory (eqn.(\ref{reduced}) below). From this it is straightforward to obtain
conserved quantities of the reduced dynamics, which can be interpreted
as the linear and angular momenta. These have some interesting features.
For example, the linear momentum turns out to be related to the mean of
the vortex positions, which is not surprising, since in a first order
dynamical system the linear momentum is often related to position.
However, the conserved linear and angular momenta in the reduced theory
were not directly related to the linear and angular momenta of the
parent field theory in \cite{Man1}. This omission is rectified here.

The conserved quantities of the parent field theory have to be derived
with care. The na\"{\i}ve canonical linear and angular momenta
are not gauge invariant; moreover, they are not conserved if the field
configuration has non-trivial topology, because of currents at
infinity. It is important to obtain gauge
invariant conservation laws for the linear and angular momenta in
the parent theory. 

The relevant conserved quantities are, in fact, known.
They have been obtained by Hassa\"{\i}ne et al. 
\cite{Has} by identifying the model
with the Jackiw--Pi model \cite{Jac} in a background field. Here,
we obtain the conservation laws from first principles using
Noether's theorem and we clarify the issues of gauge invariance and
currents at infinity. Following Papanicolaou and Tomaras \cite{Pap}, 
who studied conservation laws in a not very different model, 
we also express the
linear and angular momenta as moments of vorticity. This establishes
an analogy between our model and models of fluid vortices \cite{Bat}. 
Finally, by evaluating the linear and angular momenta 
for fields satisfying the Bogomol'nyi equations,
we will show that they coincide with the linear and
angular momenta in the reduced theory.

The outline of this paper is as follows. In Sect.2, we describe the
model, and the reduced theory of vortex dynamics obtained using the
moduli space approximation. In Sect.3, we obtain the conserved linear and 
angular momenta and express them in terms of vorticity. In Sect.4, we 
evaluate these expressions for Bogomol'nyi fields, and compare with
the conserved quantities in the reduced theory. Sect.5 contains 
our conclusions.

\section{The model}
\setcounter{equation}{0}
{(i) \sl The Schr\"{o}dinger--Chern--Simons Lagrangian}

Let $\phi$ be a complex (Higgs) scalar field representing the condensate of
the superconducting electrons and let $a_{\alpha} \ (\alpha =0,1,2)$ be
the U(1) gauge potential. We will use the subscript 0 to refer to time
and the subscripts 1,2 to refer to the two directions in space. 
The Lagrangian of the model is \cite{Man1}
\begin{equation}\label{lag}
L=T-V
\end{equation}
where the kinetic
energy
\begin{equation}\label{kin}
T=\int  \left( \gamma \frac{i}{2}(\bar{\phi}D_{0}\phi -\phi
\overline{D_{0}\phi})+ \mu (Ba_{0}+E_{2}a_{1}-E_{1}a_{2})-\gamma
a_{0}\right) d^{2}x
\end{equation}
and the potential energy
\begin{equation}\label{pot} 
V=\int \left(
\frac{1}{2}B^{2}+\frac{1}{2}\overline{D_{i}\phi}D_{i}\phi
+\frac{\lambda}{8}(1-|\phi |^{2})^{2}+a_{i}J_{i}^{T}\right) d^{2}x.
\end{equation}
Here, $\gamma, \mu$ and $\lambda$ are real constants with $\lambda$ positive, 
$D_{\alpha}\phi =(\partial_{\alpha} -ia_{\alpha})\phi $ are the
components of the covariant derivative of
$\phi$, $B=\partial_{1}a_{2}-\partial_{2}a_{1}$ is the magnetic field,
the electric field
$E_{i}=\partial_{i}a_{0}-\partial_{0}a_{i}$ and $J_{i}^{T}$ is a constant
transport current. We assume
the summation convention
in the spatial index $i = 1,2$. The Schr\"odinger term (with
coefficient $\gamma$) and the Chern--Simons term (with
coefficient $\mu$) define the kinetic energy for the scalar field and 
gauge potential. The term $\gamma a_0$, introduced by
Barashenkov and Harin \cite{BarHar}, allows the possibility of a
non-zero condensate in the ground state. The potential energy is the 
Ginzburg--Landau energy functional. Notice that the kinetic
energy contains terms with only the first power of time
derivatives. It was shown in \cite{Man1} that $L$ is Galilean
invariant. Galilean invariance determines the following role for the
transport current. Given any solution of the field equations in the
absence of a transport current, the effect of $J_{i}^{T}$ is
simply to boost the solution with a velocity
${\displaystyle{v_{i}=\frac{1}{\gamma} J_{i}^{T}}}$.
Having understood
this role of the transport current, we will henceforth neglect it.

The field equations obtained by varying respectively $\bar{\phi}$,
$a_{i}$ and $a_{0}$ are
\begin{equation}\label{e1}
i \gamma D_{0}\phi =-\frac{1}{2}D_{i}D_{i}\phi
-\frac{\lambda}{4}(1-|\phi |^{2})\phi 
\end{equation}
\begin{equation}\label{e2}
\epsilon _{ij}\partial_{j}B =J_{i}+2\mu
\epsilon _{ij}E_{j}  
\end{equation}
\begin{equation}\label{e3}
2 \mu B= \gamma (1-|\phi |^{2})
\end{equation}
where $J_{i}$ is the supercurrent defined by
\begin{equation}\label{j1}
J_{i}=-\frac{i}{2}(\bar{\phi}D_{i}\phi -\phi
{\overline{D_{i}\phi}}).
\end{equation}
Eqn.(\ref{e1}) is the gauged non-linear Schr\"{o}dinger equation,
eqn.(\ref{e2}) is
the two-dimensional version of Amp\`{e}re's law and eqn.(\ref{e3}) is a
constraint. Such a constraint appears in other
Chern--Simons type theories \cite{Jac}. It is useful to note that
this constraint is one of the Bogomol'nyi equations for vortices when
$\gamma =\mu$ \cite{Bog}. We shall assume that
${\displaystyle{1 - |\phi|^{2}}}$ and $D_{i}\phi$ decay rapidly
as $|\bf{x}|\rightarrow \infty$. Eqns.(\ref{e1})--(\ref{j1}) imply that
$D_{0}\phi$, $B$ and $E_{i}$ decay similarly.

\vskip 10pt
\noindent
{(ii) \sl Vortices}

The above model admits vortex solutions. Vortices appear whenever
there is a non-trivial winding of the map between the boundary circle
at spatial
infinity and the manifold of ground states of the scalar field, the circle
$|\phi|=1$. The relation between the winding number $N$ and the
magnetic flux is
\begin{equation}\label{vn}
\int  B \, d^{2}x = 2\pi N.
\end{equation}
$N$ can be interpreted as the vortex number.
For a vortex $N=1$ and for an anti-vortex $N=-1$. 
Henceforth, we suppose $N\geq 0$.
Later, we will define a gauge invariant vorticity
${\mathcal{V}}$, whose integral is $2\pi N$.
However, the vorticity is not simply ${\mathcal{V}}=B$.

Generally, a solution with $N$ vortices is not static, and we wish to
understand how the vortices move. However, it is by now well-known
that for special values of the couplings a large space of stable, static
$N$-vortex solutions exists, for any $N>0$ \cite{Tau}. These solutions
satisfy first order Bogomol'nyi equations, as well as the second order
Ginzburg--Landau field equations \cite{Bog}. For the model here,
Bogomol'nyi vortices occur when $\lambda =1$ and $\gamma=\mu$.
The first order Bogomol'nyi equations are
\begin{equation}\label{bog1}
(D_{1}+iD_{2})\phi =0 
\end{equation}
\begin{equation}\label{bog2}
B=\frac{1}{2}(1-|\phi |^{2}).
\end{equation}
Solutions of these equations also satisfy (\ref{e1})--(\ref{j1}), with 
$D_{0}\phi$ and $E_{i}$ vanishing. Bogomol'nyi vortices do
not exert forces on each other and this is why a static configuration of
$N$ vortices can exist. The solutions of the Bogomol'nyi
equations with winding number $N$ are uniquely specified by the 
unordered zeros of the scalar 
field, whose number, counted with multiplicity, is $N$.
These zeros are the vortex positions and we denote them $\{
{\bf{x}}^{s}: 1\leq s \leq N\}$. 
The space of solutions, called the $N$-vortex moduli space, is therefore 
topologically ${\bf{C}}^{N}/\Sigma_{N},$ where $\Sigma_{N}$ is the
permutation group on $N$ objects and the two-dimensional 
real plane is identified with the complex plane ${\bf{C}}$.
The $N$-vortex moduli space is a smooth manifold of dimension $2N$,
despite the apparent singularities where vortex positions coalesce.

Hassa\"{\i}ne et al. have recently discovered stationary Bogomol'nyi-type 
vortex solutions in this model with $\gamma \neq \mu$ \cite{Has}. The
fields satisfy (\ref{bog1}) and (\ref{e3}), and in addition $a_0$ is
proportional to $B$. One needs
$\lambda=2\gamma/\mu -\gamma^{2}/\mu^{2}$, and $\lambda$ must be positive. 
These vortices are sources for non-vanishing electric fields. 
However, we shall not consider these solutions here.

\vskip 10pt
\noindent
{(iii) \sl The reduced theory}

We consider the case where $\lambda$ is close to $1$ and $\gamma
=\mu$. We are interested in fields which remain close to $N$-vortex 
solutions of the Bogomol'ny equations, but in which the vortex
positions move slowly. In the moduli space
approximation to vortex dynamics, one can obtain a
reduced Lagrangian by simply inserting Bogomol'nyi
solutions into (\ref{lag}) and taking the vortex positions dependent on
time. 
Let us write $\phi$ as 
\begin{equation}\label{phi}
\phi =e^{\frac{1}{2}h+i\chi}. 
\end{equation}
$h$ is gauge invariant, and tends to zero at spatial infinity, but is 
singular at the vortex positions.
The first Bogomol'nyi equation (\ref{bog1}) implies
that
\begin{equation}\label{e15}
a_{i}=\frac{1}{2}\epsilon_{ij}\partial_{j}h+\partial_{i}\chi.
\end{equation}
From the second Bogomol'nyi equation (\ref{bog2}), one obtains 
\begin{equation}\label{e10}
\partial_{i}\partial_{i}h=e^{h}-1 +4\pi
\sum_{s=1}^{N}\delta^{2}({\bf{x}}-{ \bf{x}}^{s}).
\end{equation}
We assume this equation holds, even if the vortex positions
$\bf{x}^{s}$ are slowly moving.

Let us now suppose that the vortex positions are distinct, which is
the generic case. It is not difficult to allow for vortex coalescence.
$h=\log |\phi |^{2}$ has the following expansion around
the position of the $s$-th vortex 
\begin{equation}\label{hh}
h=\log
|{\bf{x}}-{\bf{x}}^{s}|^{2}+a^{s}+b_{1}^{s}(x_{1}-x_{1}^{s})+b_{2}^{s}
(x_{2}-x_{2}^{s})+
\cdots 
\end{equation} 
where $\{a^{s}, b_{i}^{s}\}$ are dependent on the positions of
the other vortices relative to ${\bf{x}}^{s}$.
The usefulness of this expansion was discovered by Samols \cite{Sam}, 
developing work of Strachan \cite{Strach}. $a^{s}$ plays no
significant role in what follows, but $b_{i}^{s}$ does. $b_{i}^{s}$ is
a measure of the lack of circular symmetry of $h$ around the vortex,
and is exponentially small if the other vortices are far away.
After various integrations, and suppression of total time derivative
terms, one obtains the manifestly gauge invariant
reduced Lagrangian \cite{Man1}
\begin{equation}\label{reduced}
L^{\mathrm{red}}=2\pi \gamma \sum_{s=1}^{N}\left(
(b_{2}^{s}+\frac{1}{2}x_{2}^{s})\dot{x}_{1}^{s}-(b_{1}^{s}+\frac{1}{2}
x_{1}^{s})\dot{x}_{2}^{s}\right) -V^{\mathrm{red}}
\end{equation}
where an overdot denotes time-derivative. 
This leads directly to equations of motion for the vortex positions. 
The potential (\ref{pot}) simplifies for solutions of the
Bogomol'nyi equations to
the integral ${\displaystyle{V^{\mathrm{red}}=\frac{\lambda-1}{8}\int
(1-|\phi|^{2})^{2}\, d^{2}x }}$, plus a constant $\pi N$, 
and this is a translationally and rotationally
invariant function of the vortex positions. Unfortunately, it appears
that $V^{\mathrm{red}}$ cannot be simplified to an explicit expression
depending only on $\{{\bf{x}}^{s}, a^{s}, b_{i}^{s}\}$. The functions
$b_{i}^{s}$ and $V^{\mathrm{red}}$ are not known explicitly as
functions of the relative positions of $N$-vortices, but
they can be calculated numerically and this has been done for 2-vortices
in \cite{Sam, Sha}.

\section{Conservation laws in the field theory}
\setcounter{equation}{0}

The linear and angular momenta for the field theory we are considering here
were obtained in \cite{Has}. Here, we give an independent derivation
from first principles.
Let $\{ \psi_{c}\} =\{ \phi,\bar{\phi},a_{0}, a_{1},a_{2} \}$,
where $c$ runs from 1 to 5. If under a variation of the fields
$\delta \psi_{c}$, the variation of the
Lagrangian density, ${\mathcal{L}}$, is
${\displaystyle{\delta {\mathcal{L}} =
\partial_{\mu}\hat{X}^{\mu}}}$, then Noether's theorem associates
a conserved current with such a variation. (Here and below, we suppress
the infinitesimal quantity multiplying such variations.) 
The Noether current, assuming the summation
convention over the index $c$, is
\begin{equation}\label{curra}
\hat{j}^{\mu}=\frac{\partial {\mathcal{L}}}{\partial
(\partial_{\mu}\psi_{c})}\delta \psi_{c}-\hat{X}^{\mu}.
\end{equation}
By Noether's theorem ${\displaystyle{\partial_{\mu}\hat{j}^{\mu}=0}}$,
and it follows that the integral of the
time component $\hat{j}^{0}$ is a conserved quantity provided that
the spatial components of the current
$\hat{j}^{1}$ and $\hat{j}^{2}$ fall off sufficiently fast at
spatial infinity.

\vskip 10pt
\noindent
{(i) \sl Energy}

The simplest conserved quantity to consider is energy.
This is related to invariance under time translation. Na\"{\i}vely, the
variations of the fields are given by their time derivatives. However,
one can supplement this by a gauge transformation with parameter
$-a_{0}$. The variations of the fields are then
\begin{equation}
\{ \delta \psi_{c}\} =\{ D_{0}\phi, \overline{D_{0}\phi}, 0, -E_{1},
-E_{2} \}
\end{equation}
and the change in ${\mathcal{L}}$ is $\displaystyle{
\delta \mathcal{L}=\partial_{\mu}X^{\mu}}$, where
\begin{equation}
X^{0}=\mathcal{L}+\gamma a_{0} -\mu a_{0}B, \ X^{1}=-\mu a_{0}E_{2}, 
\ X^{2}=\mu a_{0}E_{1}.
\end{equation}
Using (\ref{curra}), the energy density is
\begin{equation}
j^{0}=\frac{1}{2}B^{2}+\frac{1}{2}\overline{D_{i}\phi} D_{i}\phi 
+\frac{\lambda}{8}(1-|\phi |^{2})^{2}.
\end{equation}
$j^{0}$ is gauge invariant. Moreover, its integral is conserved, because the
spatial components of the currents
\begin{equation}
j^{1}=-\frac{1}{2}\overline{D_{1}\phi}{D_{0}\phi}-\frac{1}{2}D_{1}
\phi\overline{D_{0}\phi}+BE_{2}
\end{equation}
\begin{equation}
j^{2}=-\frac{1}{2}\overline{D_{2}\phi}{D_{0}\phi}-\frac{1}{2}D_{2}
\phi\overline{D_{0}\phi}-BE_{1}
\end{equation}
are gauge invariant, and hence decay rapidly at spatial infinity.
Thus, the conserved energy is $V$, as given in (\ref{pot}) (recall
that the transport current is set to zero).

\vskip 10pt
\noindent
{(ii) \sl Momentum}

Let us now find the linear momentum components, $P_{i}$, associated with
translation in the $x_{i}$-direction. First, consider translation in
the $x_{1}$-direction. The na\"{\i}ve variations of the fields are given
by the spatial derivatives in the $x_{1}$-direction. One supplements
this by a gauge transformation with parameter
$-a_{1}$. The variations of the fields are then 
\begin{equation}
\{ \delta
\psi_{c}\} =\{ D_{1}\phi, \overline{D_{1}\phi}, E_{1}, 0, B\}
\end{equation}
and the change
in $\mathcal{L}$ is ${\displaystyle{\delta
{\mathcal{L}} =\partial_{\mu}X'^{\mu}}}$, where
\begin{equation}
X'^{0}=-\mu a_{1}B+\gamma a_{1}, \
X'^{1}={\mathcal{L}}-\mu a_{1}E_{2}, \
X'^{2}=\mu a_{1}E_{1}. 
\end{equation}
The density of the linear momentum in the
$x_{1}$-direction, calculated using (\ref{curra}), is
\begin{equation}\label{poti}
j'^{0}=-\gamma (
J_{1}+a_{1}).
\end{equation}
Notice that $j'^{0}$ is not gauge invariant. Moreover,
the spatial components of the currents are
\begin{equation}
j'^{1}=-\gamma \frac{i}{2}(\bar{\phi}D_{0}\phi-\phi
\overline{D_{0}\phi})-\frac{1}{2}B^{2}-\frac{1}{2}|D_{1}\phi|^{2}
+\frac{1}{2}|D_{2}\phi|^{2}+\frac{\lambda}{8}(1-|\phi|^{2})^{2} +\gamma a_{0} 
\end{equation}
and
\begin{equation}
j'^{2}=-\frac{1}{2}(\overline{D_{2}\phi} D_{1}\phi + D_{2}\phi
\overline{D_{1}\phi}).
\end{equation} 
$j'^{1}$ is not gauge invariant, and hence does not fall off sufficiently
fast at infinity for the integral of $j'^{0}$ to be conserved. 
The remedy for both problems is to note that
$X'^{\mu}$ is not uniquely defined, but can be altered by
adding total derivative terms. One chooses an improved
$\tilde{X} ^{\mu}$, with
${\displaystyle{\partial_{\mu}X'^{\mu}=\partial_{\mu}\tilde{X}'^{\mu}}}$,
in
such a way that the resulting current is gauge invariant. 
$\tilde{X}'^{\mu}$ can be taken as
\begin{equation}
\tilde{X}'^{0}=X'^{0}+\gamma \partial_{1}(
x_{2}a_{2})-\gamma \partial_{2}( x_{2}a_{1})
\end{equation}
\begin{equation}
\tilde{X}'^{1}=X'^{1}+\gamma \partial_{2}( x_{2}a_{0})
-\gamma \partial_{0}(x_{2}a_{2})
\end{equation}
\begin{equation}
\tilde{X}'^{2}=X'^{2}+\gamma \partial_{0}(
x_{2}a_{1})-\gamma \partial_{1}( x_{2}a_{0}).
\end{equation}
Using
$\tilde{X}'^{\mu}$, the improved density of the linear momentum in the
$x_{1}$-direction is
\begin{equation}
\tilde{j}'^{0}=-\gamma(J_{1}+x_{2}B)
\end{equation}
and the spatial components of the current are 
\begin{equation}
\tilde{j}'^{1}=-\gamma \frac{i}{2}(\bar{\phi}D_{0}\phi-\phi
\overline{D_{0}\phi})-\frac{1}{2}B^{2}-\frac{1}{2}|D_{1}\phi|^{2}
+\frac{1}{2}|D_{2}\phi|^{2}+\frac{\lambda}{8}(1-|\phi|^{2})^{2}
-\gamma x_{2}E_{2}
\end{equation}
\begin{equation}
\tilde{j}'^{2}=-\frac{1}{2}(\overline{D_{2}\phi} D_{1}\phi +D_{2}\phi
\overline{D_{1}\phi})+\gamma x_{2}E_{1}.
\end{equation}
Clearly, $\tilde{j}'^{1}$ and $\tilde{j}'^{2}$ are now
gauge invariant
and fall off sufficiently fast at spatial infinity. Similarly, one can
consider translations
in the $x_{2}$-direction. One concludes that the 
conserved linear momentum is
\begin{equation}
P_{i}=-\int \tilde{j}'^{0} d^{2}x= \gamma \int
(J_{i}+\epsilon_{ij}x_{j}B) \, d^{2}x.
\end{equation}
(The choice of minus sign, here and in (\ref{M}) below, is deliberate,
and ensures agreement between the conservation laws in the field
theory and those in the reduced theory. It is made because the 
field variations are due to passive coordinate variations, whereas later we
will be actively varying the vortex positions in the reduced theory.)

\vskip 10pt
\noindent
{(iii) \sl Angular momentum}

Now, let us obtain the conservation law for angular momentum, $M$, by
considering the generator of rotations, $\epsilon_{ij}x_{i}\partial_{j}$,
combined with a gauge transformation with parameter
$-\epsilon_{ij}x_{i}a_{j}$. Here, care is needed to evaluate the Lie 
derivatives correctly on the scalar field and gauge potential. 
The variations of the fields are
\begin{equation}
\{ \delta \psi_{c}\} =\{ \epsilon_{ij}x_{i}D_{j}\phi,
\epsilon_{ij}x_{i}\overline{ D_{j}\phi},\epsilon_{ij}x_{i}E_{j},
-x_{1}B, -x_{2}B \} .
\end{equation}
The change in $\mathcal{L}$ is ${\displaystyle{\delta
\mathcal{L}=\partial_{\mu} X''^{\mu}}}$, where
\begin{equation}
X''^{0}=-\epsilon_{ij}x_{i}a_{j}(\mu B-\gamma ), \
X''^{1}=-x_{2}\mathcal{L}-\mu \epsilon_{ij}x_{i}a_{j}E_{2}, \
X''^{2}=x_{1}\mathcal{L} +\mu \epsilon_{ij}x_{i}a_{j}E_{1}.
\end{equation}     
The angular momentum density obtained using (\ref{curra}) is
\begin{equation}
j''^{0}=-\gamma \epsilon_{ij}x_{i}(J_{j}+a_{j}).
\end{equation}
Neither this density nor the spatial components of the current are 
gauge invariant, nor do they fall off sufficiently fast
at spatial infinity. Again, the remedy
is to find an improved $\tilde{X}''^{\mu}$, with
${\displaystyle{\partial_{\mu}X''^{\mu}=\partial_{\mu}\tilde{X}''^{\mu}}}$.
One may take
\begin{equation}
\tilde{X}''^{0}=X''^{0}-\gamma \partial_{1}(a_{2}r^{2}/2)+\gamma
\partial_{2}(a_{1}r^{2}/2)
\end{equation}
\begin{equation}
\tilde{X}''^{1}=X''^{1}-\gamma
\partial_{2}(a_{0}r^{2}/2)+\gamma \partial_{0}(a_{2}r^{2}/2)
\end{equation}
\begin{equation}
\tilde{X}''^{2}=X''^{2}-\gamma \partial_{0}(a_{1}r^{2}/2)+\gamma
\partial_{1}(a_{0}r^{2}/2).
\end{equation}
Then, the improved angular momentum density is
\begin{equation}
\tilde{j}''^{0}=-\gamma (\epsilon_{ij}x_{i}J_{j}-\frac{1}{2}r^{2}B)
\end{equation}
which is clearly gauge invariant. Likewise, $\tilde{j}''^{1}$ and
$\tilde{j}''^{2}$ are gauge invariant and do fall off sufficiently fast.
The conserved angular momentum is therefore
\begin{equation}\label{M}
M=-\int \tilde{j}''^{0} d^{2}x= \gamma \int \left(
\epsilon_{ij}x_{i}J_{j}- \frac{1}{2}r^{2}B\right)
\,  d^{2}x.
\end{equation}

\vskip 10pt
\noindent
{(iv) \sl Vorticity}

Let us now define the vorticity 
\begin{equation}\label{v11}
{\mathcal{V}}=\epsilon_{ij}\partial_{i}J_{j}+B.
\end{equation}
Substituting for $J_{i}$, using (\ref{j1}), the vorticity can be
written as
\begin{equation}\label{v111}
\mathcal{V}=-i\epsilon_{ij}\overline{D_{i}\phi}D_{j}\phi
+B(1-|\phi|^{2}),
\end{equation}
which is the definition in \cite{Pap}, and is a gauge
invariant generalisation of the notion of vorticity discussed in
\cite{Pap1}. In the sector with vortex number $N$,
\begin{equation}\label{vo}
\int \mathcal{V}\, d^{2}x =2\pi N,
\end{equation}
using (\ref{v11}) and Stokes' theorem. 
Integrating by parts, one may express the linear and angular
momenta as the following moments of the vorticity
\begin{equation}\label{puii1}
P_{i}=\gamma \epsilon_{ij} \int x_{j}\mathcal{V}\, d^{2}x
\end{equation}
and 
\begin{equation}\label{puii2}
M=-\frac{\gamma}{2} \int r^{2}\mathcal{V}\, d^{2}x.
\end{equation}
We noted in the introduction that in a first order dynamical system the
linear momentum can often be taken as a measure of position. 
The formula (\ref{puii1}) shows that this idea applies here.
The components of momentum are proportional
to the components of the centre of vorticity.
Eqns. (\ref{puii1}) and (\ref{vo}) imply that 
$R_{i}=-{\displaystyle{\frac{1}{2\pi N \gamma}}}\epsilon_{ij}P_{j}$ is
the centre of vorticity \cite{Pap}, and it does not move. 

The conservation of the angular momentum (\ref{puii2}) shows that no
net vorticity can escape to infinity, assuming that singularities
do not form, and therefore a vortex cannot escape to infinity, unless 
accompanied by an anti-vortex.

The vorticity has the following interesting property
for Bogomol'nyi vortices. Substituting (\ref{bog2})
in (\ref{pot}),
the energy density for Bogomol'nyi vortices is seen to be
\begin{equation}
{\mathcal{E}}^{\mathrm{Bog}}=\frac{1}{2}|D_{i}\phi |^{2}+B^{2}.
\end{equation}
On the other hand, using both Bogomol'nyi equations,
the vorticity (\ref{v111}) can be rewritten as
\begin{equation}
\mathcal{V}=|D_{i}\phi |^{2}+2B^{2}.
\end{equation}
Thus, for Bogomol'nyi vortices
${\displaystyle{{\mathcal{V}}=2{\mathcal{E}}^{\mathrm{Bog}}}}$.

\section{Conservation laws in the reduced dynamics}
\setcounter{equation}{0}
{(i) \sl The conserved quantities}

Conservation laws of the reduced dynamics can be obtained directly
from ${L^{\mathrm{red}}}$, eqn.(\ref{reduced}). (Note that the discussion
of conservation laws at the end of ref.\cite{Man1} is slightly wrong.)
In general, a variation $\delta x_{i}^{s} =
\xi_{i}^{s}$ is a symmetry if $\delta L^{\mathrm{red}} = 
\frac{d}{dt}X$ for some $X$.  
Noether's theorem states that 
\begin{equation}
\sum_{s=1}^{N}\frac{\partial {L^{\mathrm{red}}}} {\partial
\dot{x}_{i}^{s}} \xi_{i}^{s} - X
\end{equation}
is then conserved. 

A translation of all the vortex positions in the $x_{1}$-direction is
a symmetry. Here, for all $s$,
\begin{equation}
\delta x_{1}^{s} = 1 , \ \delta x_{2}^{s} = 0,
\end{equation}
and $b_{i}^{s}$ and $V^{\mathrm{red}}$ are invariant. One finds that 
${\displaystyle{ X=-\pi \gamma \sum_{s=1}^{N} x_{2}^{s}}}$, and the
conserved component of momentum is
\begin{equation}
P_{1}^{\mathrm{red}}=2\pi \gamma 
\sum_{s=1}^{N}(b_{2}^{s}+x_{2}^{s}).
\end{equation}
Similarly, translation in the $x_{2}$-direction is a symmetry, and 
\begin{equation}
P_{2}^{\mathrm{red}}=-2\pi \gamma 
\sum_{s=1}^{N}(b_{1}^{s}+x_{1}^{s})
\end{equation}
is conserved. It was shown by Samols that 
${\displaystyle{\sum_{s=1}^{N}b_{i}^{s}=0}}$. The 
linear momentum in the reduced dynamics is therefore simply
\begin{equation}\label{pr}
P_{i}^{\mathrm{red}}=2\pi \gamma \epsilon_{ij}\sum_{s=1}^{N}x_{j}^{s},
\end{equation}
and directly related to the mean of the vortex positions. Conservation
of momentum implies that the vortices circulate about their fixed mean
position.

There is also symmetry under a rotation, where, for all $s$
\begin{equation}
\delta x_{1}^{s} = -x_{2}^{s} , \
\delta x_{2}^{s} = x_{1}^{s}.
\end{equation}
$V^{\mathrm{red}}$ is invariant, but the rotation leads to the
variations
\begin{equation}
\delta b_{1}^{s} = -b_{2}^{s} , \
\delta b_{2}^{s} = b_{1}^{s}.
\end{equation}
It follows that $L^{\mathrm{red}}$ is strictly invariant, with $X=0$, so 
one has the following conserved angular momentum in the reduced dynamics 
\begin{equation}\label{lr}
M^{\mathrm{red}}=-2\pi \gamma \sum_{s=1}^{N}\left(
\frac{1}{2}x_{i}^{s}x_{i}^{s}+b_{i}^{s}x_{i}^{s}\right).
\end{equation}
This conservation law probably implies that no vortex can escape to
infinity, but we do not have enough information on the functions 
$b_{i}^{s}$ to prove this. If the $s$-th vortex went to infinity, 
$x_{i}^{s}x_{i}^{s}$ would diverge while $b_{i}^{s}x_{i}^{s}$ would tend to
zero. This could be compensated by singularities in the values of  
$b_{i}^{s'} \ (s' \neq s)$. Now such singularities do occur as
vortices coalesce \cite{Sam}, but they appear (at least for two vortices) to
cancel in ${\displaystyle{\sum_{s=1}^{N}b_{i}^{s}x_{i}^{s}}}$, leaving
a finite result.

\vskip 10pt
\noindent
{\sl (ii) Comparison with the field theory}

We compare the conserved quantities in our field theory
with the corresponding conserved quantities obtained directly from the
reduced Lagrangian, $L^{\mathrm{red}}$, by assuming the fields
satisfy the Bogomol'nyi equations at all times, possibly with
time-varying vortex positions. This is sensible if $\lambda$ is close to
$1$ and $\mu = \gamma$. 

First of all, the conserved
energy is $E = \pi N + V^{\mathrm{red}}$. This is consistent with
the reduced dynamics, where the Hamiltonian is simply
$V^{\mathrm{red}}$ (the constant $\pi N$ is dropped), 
and $V^{\mathrm{red}}$ is conserved.

The main task is to evaluate the expressions (\ref{puii1}) and
(\ref{puii2}) for linear and angular momentum. Using (\ref{phi}) 
and (\ref{e15}),
it can be shown that for solutions of the Bogomol'nyi equations
\begin{equation}\label{jos}
J_{i}=-\frac{1}{2}\epsilon_{ij}\partial_{j}h \, e^{h}, \; \; \; 
B=-\frac{1}{2}\partial_{i}\partial_{i}h.
\end{equation}
From (\ref{v11}), the vorticity ${\mathcal{V}}$ becomes
\begin{equation}\label{vori1}
{\mathcal{V}}=\frac{1}{2}\partial_{i}\partial_{i}(e^{h}-h).
\end{equation}
Another expression for ${\mathcal{V}}$ is
\begin{equation}\label{vori2}
{\mathcal{V}}=\frac{1}{2}\partial_{i}(\partial_{i}h(e^{h}-1))=
\frac{1}{2}\partial_{i}
(\partial_{i}h \partial_{j}\partial_{j}h),
\end{equation}
where use has been made of (\ref{e10}) and temporarily we ignore the
logarithmic singularities of $h$. Note that ${\mathcal{V}}$ is a 
smooth function despite the singularities of $h$.

In what follows we will again suppose that $\phi$ has $N$ simple zeros.
In order to carry out the integrals involving moments of ${\mathcal{V}}$
let us remove, from ${\bf{R}}^{2}$, $N$
discs of small radius $\epsilon$ centred at the vortex positions,
and call the resulting region ${\bf{R}}_{0}^{2}$. As ${\mathcal{V}}$ is a
smooth function, integrations over the discs will be
$O(\epsilon^{2})$ or smaller, and hence can be neglected in the limit
$\epsilon \rightarrow 0$. Thus, in the following, the effective region
of integration is ${\bf{R}}_{0}^{2}$, and the 
singularities of $h$ may
be ignored in the formulae (\ref{vori1}) and (\ref{vori2}) for ${\mathcal{V}}$.

Let $C^{s}$, where $s$ runs from 1 to $N$,
denote the boundary of the disc around the $s$-th vortex position 
${\bf{x}}^{s}$
and let $C^{0}$ denote the boundary circle at spatial
infinity. Further, let $\theta^{s}$ be the polar angle relative to 
${\bf{x}}^{s}$ with $\theta^{s}=0$ in the positive
$x_{1}$-direction. Then, the outward unit normal along $C^{s}$ is
${\bf{n}}^{s}=(\cos \theta^{s}, \sin
\theta^{s}) $ and the coordinates of points on $C^{s}$ can be written 
as $x_{i}=x_{i}^{s} +\epsilon n_{i}^{s}$. The differential line element on
$C^{s}$ is $dl=\epsilon d\theta^{s}$.

Now, using (\ref{vori1}), and remembering the discussion above,
the linear momentum (\ref{puii1}) can be
written with $O(\epsilon^{2})$ error as
\begin{equation}
P_{i}=\frac{\gamma}{2}\epsilon_{ij} \int_{{\bf{R}}_{0}^{2}}\left(
x_{j}\partial_{k}\partial_{k}(e^{h}-h-1)\right) d^{2}x.
\end{equation}
The replacement of ${\displaystyle{(e^{h}-h)}}$ by
${\displaystyle{(e^{h}-h-1)}}$ is convenient at this stage.
Using Green's lemma in two dimensions,
\begin{equation}
P_{i}=-\frac{\gamma}{2}\epsilon_{ij} 
\sum_{s=1}^{N} \int_{C^{s}}\left(
x_{j}\partial_{k}(e^{h}-h-1)-(e^{h}-h-1)\partial_{k}x_{j}\right)
n_{k}^{s}\, dl
\end{equation}
\begin{equation}
=-\frac{\gamma}{2}\epsilon_{ij} 
\sum_{s=1}^{N} \int_{C^{s}}\left(
x_{j}\partial_{k}h(e^{h}-1)n_{k}^{s}-(e^{h}-h-1)n_{j}^{s}\right)
\, dl.
\end{equation}
There is no contribution from $C^{0}$, the circle at infinity, as
${\displaystyle{e^{h}-h-1}}$ vanishes there.
In calculating the 
integrals along $C^{s}$ we will ignore terms which are 
$O(\epsilon)$ or smaller. On $C^{s}$, one finds from
(\ref{hh}) that
\begin{equation}
\partial_{k}h=\frac{2n_{k}^{s}}{\epsilon}+b_{k}^{s} + \dots,
\end{equation}
and $e^{h}=O(\epsilon^{2})$.
Therefore
\begin{equation}
P_{i}=-\frac{\gamma}{2}\epsilon_{ij}
\sum_{s=1}^{N}\int_{C^{s}}\left( -(x_{j}^{s}+\epsilon
n_{j}^{s})(\frac{2n_{k}^{s}}{\epsilon}+b_{k}^{s}) n_{k}^{s}+(\log
\epsilon^{2}+a_{s}+1)n_{j}^{s}\right) dl.
\end{equation}
Noting that ${\displaystyle{\int_{C^{s}}n_{j}^{s}\, dl=0}}$ and
${\displaystyle{\int_{C^{s}}n_{j}^{s}n_{k}^{s}dl=\pi\epsilon
\delta_{jk}}}$,
one concludes that 
\begin{equation}
P_{i}=2\pi \gamma \, \epsilon_{ij}\sum_{s=1}^{N}x_{j}^{s},
\end{equation}
which is the same as the expression (\ref{pr}), derived in the reduced
dynamics.

Before proceeding to compute the angular momentum $M$
we note the following useful identity
\begin{equation}
r^{2}{\mathcal{V}}=\frac{1}{2}r^{2}\partial_{i}
(\partial_{i}h\partial_{j}
\partial_{j}h)=\frac{1}{2}\partial_{i}q_{i},
\end{equation}
where
\begin{equation}
q_{i}=r^{2}\partial_{i}h \partial_{j}\partial_{j}h-
2x_{j}\partial_{j}h\partial_{i}h+x_{i}\partial_{j}h\partial_{j}h.
\end{equation}
As $r^{2}{\mathcal{V}}$ is a smooth function, we follow the same
procedure as
in evaluating $P_{i}$, namely, remove $N$ small discs centred at
the positions of vortices. With $O(\epsilon^{2})$ error, 
\begin{equation}
M=-\frac{\gamma}{2} \int r^{2}{\mathcal{V}} d^{2}x=
-\frac{\gamma}{4}\int_{{\bf{R}}^{2}_{0}}
\partial_{i}q_{i} d^{2}x = \frac{\gamma}{4}\sum_{s=1}^{N}
\int_{C^{s}}q_{i}n_{i}^{s} \, dl.
\end{equation}
Again there is no contribution coming from $C^{0}$, the circle at infinity,
as $\partial_{i}h$ vanishes there.
We rewrite $M$ as
\begin{equation}
M=\frac{\gamma}{4} \sum_{s=1}^{N}(I_{1}^{s}-2I_{2}^{s}+I_{3}^{s}),
\end{equation}
where
\begin{equation}
I_{1}^{s}=\int_{C^{s}} \left( r^{2}\partial_{i}h \partial_{j}
\partial_{j}h \right) n_{i}^{s} \, dl,
\end{equation}
\begin{equation}
I_{2}^{s}=\int_{C^{s}}\left(
x_{j}\partial_{j}h \partial_{i}h \right) n_{i}^{s} \, dl,
\end{equation}
and
\begin{equation}
I_{3}^{s}=\int_{C^{s}}\left(
x_{i} \partial_{j}h \partial_{j}h \right) n_{i}^{s} \, dl.
\end{equation}
Noting from (\ref{hh}) that on $C^{s}$, $\partial_{i}\partial_{i}h=-1
+ O(\epsilon^{2})$, we obtain
\begin{equation}
I_{1}^{s}=\int_{C^{s}}(x_{k}^{s}x_{k}^{s}+2\epsilon
n_{k}^{s}x_{k}^{s})(\frac{2}{\epsilon}
+b_{i}^{s}n_{i}^{s})(-1) \, dl=-4\pi
x_{k}^{s}x_{k}^{s} 
\end{equation}
where, as usual, terms of $O(\epsilon)$ or smaller have been neglected.
Similarly,
\begin{equation}
I_{2}^{s}=\int_{C^{s}}(\frac{2}{\epsilon}x_{j}^{s}n_{j}^{s} 
+2+x_{j}^{s}b_{j}^{s})
(\frac{2}{\epsilon} +b_{i}^{s}n_{i}^{s}) \, dl =8\pi+6\pi
b_{j}^{s}x_{j}^{s},
\end{equation}
and
\begin{equation}
I_{3}^{s}=\int_{C^{s}
}(x_{i}^{s}n_{i}^{s}+\epsilon)(\frac{4}{\epsilon^{2}} 
+ \frac{4}{\epsilon}b_{j}^{s}n_{j}^{s} +b_{j}^{s}b_{j}^{s}) \,
dl =8\pi + 4\pi b_{j}^{s}x_{j}^{s}.
\end{equation}
Thus, putting all the above integrals together,
\begin{equation}\label{m3}
M=-2\pi \gamma \sum_{s=1}^{N}\left(
\frac{1}{2}x_{i}^{s}x_{i}^{s}+b_{i}^{s}x_{i}^{s}+1\right),
\end{equation}
which apart from a constant additive term agrees with (\ref{lr}).

The constant term which appears in (\ref{m3}) has the following
meaning. A single vortex situated at the origin has no linear
momentum. However, its total angular momentum is $-2\pi \gamma$. 
Thus, associated with each vortex there is a net spin. 

\vskip 10pt
\noindent
{\sl (iii) Contribution of the supercurrent to the momenta}

It is of some interest to separate the contributions of the
supercurrent and the magnetic field to the linear and angular momenta,
for fields satisfying
the Bogomol'nyi equations. We note that the vorticity
${\mathcal{ V}}$
can be written as
\begin{equation}
{\mathcal{V}}={\mathcal{V}}_{J}+B
\end{equation}
where the contribution due to the supercurrent is
\begin{equation}
{\mathcal{V}}_{J}=\epsilon_{ij}\partial_{i}J_{j}.
\end{equation}
Note that the integral of ${\mathcal{V}}_{J}$ over the plane is zero,
so the total vorticity, or vortex number, is entirely due to the 
magnetic field.
>From (\ref{vori1}),
\begin{equation}\label{sup}
{\mathcal{V}}_{J}
=\frac{1}{2}\partial_{i}\partial_{i}(e^{h})=\frac{1}{2}\partial_{i}
\partial_{i}
(\partial_{j}\partial_{j}h).
\end{equation}
It is not difficult to show that for $k=1,2$
\begin{equation}
\int x_{k}{\mathcal{V}}_{J}\, d^{2}x =0.
\end{equation}
The supercurrent therefore makes no contribution
to the linear momentum, and
\begin{equation}
P_{i}=\gamma \, \epsilon_{ij}\int x_{j}B \, d^{2}x.
\end{equation}
Further, it can be shown that
\begin{equation}
\int (x_{k})^{2}{\mathcal{V}}_{J} \, d^{2}x= -4\pi N.
\end{equation}
Hence, the supercurrent contribution to the angular momentum $M$ is
\begin{equation}
M_{J}=-\frac{\gamma}{2}\int r^{2}{\mathcal{V}}_{J}\, d^{2}x=4\pi
\gamma N ,
\end{equation}
and the contribution due to $B$ is therefore
\begin{equation}
M_{B}=-\frac{\gamma }{2}\int r^{2}B \, d^{2}x =-2\pi \gamma
\sum_{s=1}^{N}\left( \frac{1}{2}x_{i}^{s}x_{i}^{s} +b_{i}^{s}x_{i}^{s}+3
 \right).
\end{equation}
Thus, in the reduced dynamics, ${\mathcal{V}}_{J}$ 
contributes just a constant to the angular momentum.

The calculations above make it interesting to compute the third and 
the fourth moments of ${\mathcal{V}}_{J}$, for fields satisfying
the Bogomol'nyi equations. 
One can show, using (\ref{sup}), that
\begin{equation}
\int (x_{k})^{3}\mathcal{V}_{J}\, d^{2}x=-12\pi \sum_{s=1}^{N}x_{k}^{s}.
\end{equation}
To obtain the fourth moment it is necessary to compute the
integral of $h$ over the plane. This integral is finite despite the
logarithmic singularities of $h$. Using (\ref{e10}), the vorticity 
(\ref{vori1}) can be written as
\begin{equation}
\mathcal{V}=\frac{1}{2}\partial_{i}\partial_{i}\partial_{j}\partial_{j}h-
\frac{1}{2}\partial_{i}\partial_{i}h.
\end{equation}
Then,
\begin{equation}
M=-\frac{\gamma}{2}\int r^{2}\mathcal{V}\, d^{2}x=\gamma \int h\, d^{2}x
-\pi \gamma
\sum_{s=1}^{N}x_{i}^{s}x_{i}^{s}+4\pi \gamma N.
\end{equation}
Equating this with (\ref{m3}), we find the noteworthy result
\begin{equation}\label{h}
\int h \, d^{2}x=-2\pi\sum_{s=1}^{N}\left( b_{i}^{s}x_{i}^{s}+3 \right).
\end{equation}
Using (\ref{sup}), and Green's lemma, we find
\begin{equation}
\int (x_{k})^{4}\mathcal{V}_{J}\, d^{2}x=6\int (x_{k})^{2}
\partial_{j}\partial_{j}h \, d^{2}x
=6\int \left(
\partial_{j}((x_{k})^{2} \partial_{j}h)-2
\partial_{k}(x_{k}h) +2 h \right) d^{2}x,
\end{equation}
where there is no summation over the index $k$. Integrating again, 
and using (\ref{h}), we conclude that
\begin{equation}
\int (x_{k})^{4}\mathcal{V}_{J}\, d^{2}x=-24\pi
\sum_{s=1}^{N}((x_{k}^{s})^{2}+b_{i}^{s}x_{i}^{s}+3).
\end{equation}

\section{Conclusions}

In this paper, we have succeeded in obtaining
conservation laws from first principles for the
linear and angular momenta in the Schr\"{o}dinger--Chern--Simons
dynamical model of vortices as proposed in \cite{Man1}. Quite similar 
to fluid vortices, the linear and angular momenta can be expressed as 
low moments of a suitably defined vorticity. Our
expressions agree with those in \cite{Has} in the absence of any
transport current. The conserved quantities in the presence of a
transport current are those that follow using the Galilean invariance
of the dynamics.

For a range of values of the couplings, vortex dynamics in the model  
reduces, approximately, to motion in the moduli space of 
Bogomol'nyi vortices. The expressions for the linear and angular
momenta in the reduced theory have been shown to agree with those
obtained by evaluating the linear and angular momenta in the 
parent field theory, assuming the fields satisfy the Bogomol'nyi equations.
This agreement was not inevitable, and supports the use of the moduli 
space approximation. 
Various integrals involving the vorticity have been
evaluated explicitly to make the comparison possible. One consequence
of the calculations is that each vortex has a constant net spin.
Our results can probably be extended somewhat, to a larger range of
couplings, by exploiting the electrically excited vortex solutions of
Hassa\"{\i}ne et al. \cite{Has}. 

The conservation of linear momentum implies that the centre of
vorticity, which becomes the mean of the vortex positions in the
reduced theory, does not move. Conservation of angular momentum probably
implies that net vorticity cannot escape to infinity, and hence no
vortex can escape to infinity in the reduced dynamics, but this result
is not proved. Recently, there have been numerical 
studies of vortices in the model considered here
\cite{Str}. The conservation laws should 
be a useful guide to the accuracy of such numerical studies.

\vspace{.3cm}
{\centerline{\bf{Acknowledgements}}}
\vspace{.25cm}

NSM is grateful to N. Papanicolaou and P. Horv\'{a}thy, and their
collaborators, for very helpful discussions and hospitality.


\newpage
\begin{thebibliography}{18}

\bibitem{Ait} Aitchison I J R, Ao P, Thouless D L and Zhu X -M 1995
{\sl Phys. Rev.} B {\bf{51}} 6531 

\bibitem{BarHar} Barashenkov I V and Harin A O 1994 {\sl
Phys. Rev. Lett.} {\bf{72}} 1575

\bibitem{Bat} Batchelor G K 1967 {\sl An introduction to fluid
dynamics} \S 7.3 (Cambridge: Cambridge University Press)

\bibitem{Bog} Bogomol'nyi E B 1976 
{\sl Sov. J. Nucl. Phys.} {\bf{24}} 449 

\bibitem{Has} Hassa\"{\i}ne M, Horv\'{a}thy P A and Yera J -C 1998
{\sl Ann. Phys.} {\bf{263}} 276

\bibitem{Jac} Jackiw R and Pi S -Y 1990 {\sl Phys. Rev.} D {\bf{42}} 3500 

\bibitem{Man2} Manton N S 1982
{\sl Phys. Lett.} B {\bf{110}} 54 

\bibitem{Man1} Manton N S 1997 {\sl Ann. Phys.} {\bf{256}} 114 

\bibitem{Pap} Papanicolaou N and Tomaras T N 1993
{\sl Phys. Lett.} A {\bf{179}} 33

\bibitem{Pap1} Papanicolaou N and Tomaras T N 1991 
{\sl Nucl. Phys.} B {\bf{360}} 425;
Komineas S and Papanicolaou N 1998 {\sl Nonlinearity} {\bf{11}} 265 

\bibitem{Sam} Samols T M 1992 {\sl Commun. Math. Phys.} {\bf{145}} 149 

\bibitem{Sha} Shah P A 1994 {\sl Nucl. Phys.} B {\bf{429}} 259 

\bibitem{Sto} Stone M 1995 {\sl Int. J. Mod. Phys.} B {\bf{9}} 1359 

\bibitem{Strach} Strachan I A B 1992 {\sl J. Math. Phys.} {\bf{33}} 102

\bibitem{Str} Stratopoulos G 1998 unpublished

\bibitem{Stu} Stuart D 1994 {\sl Commun. Math. Phys.} {\bf{159}} 51

\bibitem{Tau} Taubes C H 1980 {\sl Commun. Math. Phys.} {\bf{72}}
277 










\end{thebibliography}
\end{document}